\documentclass[11pt]{article}
\usepackage{hyperref}
\begin{document}
\title{Superheated water drops in hot oil
}
\author{Enrique Soto, Roberto Zenit and Andrew Belmonte* \\
\\\vspace{6pt} Universidad Nacional Aut\'onoma de M\'exico, Ciudad Universitaria, M\'exico \\ Penn State University, University Park, PA 16802, USA}
\maketitle
\begin{abstract}

Drops of water at room temperature were released in hot oil, which had a temperature higher than that of the boiling point of water. Initially, the drop temperature increases slowly mainly due to heat transfer diffusion; convective heat transfer is small because the motion takes place at a small Reynolds number. Once the drop reaches the bottom of the container, it sticks to the surface with a certain contact angle. Then, a part of the drop vaporizes: the nucleation point may appear at the wall, the interface or the bulk of the drop. The vapor expands inside the drop and deforms its interface. The way in which the vapor expands, either smooth or violent (Brennen,1995, 2002), depends on the location of the nucleation point and oil temperature. Furthermore, for temperatures close to the boiling point of water, the drops are stable (overheated); the vaporization does not occur spontaneously but it may be triggered with an external perturbation (Ghiaasiaan,2008). In this case the growth of the vapor bubble is rather violent. Many visualization for different conditions will be shown ( \href{http://hdl.handle.net/1813/14113}{Video} ) and predictions of the growth rate will be discussed.

\end{abstract}
\begin{enumerate}
\item Brennen, C.E \emph{Fission of collapsing cavitation bubbles}, J Fluid Mech , \textbf{472}, 153-166 (2002).

\item S. Mostafa Ghiaasiaan \emph{Two-Phase Flow, Boiling and Condensation}, Cambridge University Press, New York, (2008).
\item Brennen, C.E \emph{Cavitation and Bubble Dynamics}, Oxford
University Press, Chap \textbf{3} (1995).

\end{enumerate}

\end{document}